# Probing Lee-Yang zeros and coherence sudden death

Bo-Bo Wei[1,], & Ren-Bao Liu[1,2,*]

1. *Department of Physics and Centre for Quantum Coherence, The Chinese University of Hong Kong, Hong Kong, China*
2. *Institute of Theoretical Physics, The Chinese University of Hong Kong, Hong Kong, China*

*\* Corresponding author. rbliu@phy.cuhk.edu.hk*

**As a foundation of statistical physics, Lee and Yang in 1952 proved that the partition functions of thermal systems can be zero at certain points (called Lee-Yang zeros) on the complex plane of temperature [1,2]. In the thermodynamic limit, the Lee-Yang zeros approach to real numbers at the critical temperature [1]. However, the imaginary Lee-Yang zeros have not been regarded as experimentally observable since they occur at imaginary field or temperature, which are unphysical. Here we show that the coherence of a probe spin weakly coupled to a many-body system presents zeros as a function of time that are one-to-one mapped to the Lee-Yang zeros of the many-body system. In the thermodynamic limit, of which the Lee-Yang zeros form a continuum, the probe spin coherence presents a sudden death at the edge singularities of the Lee-Yang zeros [3,4]. By measuring the probe spin coherence, one can directly reconstruct the partition function of a many-body system. These discoveries establish a profound relation between two most fundamental quantities in the physical world, time and temperature, and also provide a universal approach to studying interacting many-body systems through measuring coherence of only one probe spin (or one qubit in quantum computing).**

In 1952 Yang and Lee [1] proposed to analyze phase transitions by studying partition function zeros, termed as Lee-Yang zeros, in the complex fugacity plane. They proved the famous unit-circle theorem [2] which we briefly describe below. Consider a general Ising model with ferromagnetic interaction under a magnetic field $h$, with Hamiltonian

$$H(h) = -\sum_{i,j} J_{ij}\sigma_i\sigma_j - h\sum_j \sigma_j, \tag{1}$$

where the spins $\sigma_j$ take values $\pm 1$ and $J_{ij} \geq 0$. The partition function of $N$ spins at temperature $T$ can be written into an $N$-th order polynomial of $z \equiv \exp(-2\beta h)$ as $Z(\beta,h) = \text{Tr}\left[\exp(-\beta H)\right] = e^{\beta Nh}\sum_{n=0}^{N} p_n z^n$, where $\beta \equiv 1/T$ is the inverse temperature (Boltzmann and Planck constants taken as unity) and $p_n$ is the partition function in zero magnetic field under the constraint that $n$ spins are at state $-1$. The parameter $z$ can be regarded as the scaled field or scaled inverse temperature. Lee and Yang [2] proved that the $N$ zeros of the partition function all lie on a unit circle in the complex plane of $z$ (corresponding to complex $h$ for fixed temperature or complex temperature for fixed field). We denote the zeros as $z_n \equiv \exp(i\theta_n)$, where $n = 1, 2, \cdots, N$. If the Lee-Yang zeros are determined by certain methods, the partition function can be readily reconstructed as $Z = e^{\beta Nh}\prod_{n=1}^{N}(z - z_n)$. Above the critical temperature $T_c$, the partition function is nonzero throughout the neighborhood of the real axis of $z$ within the gap determined by two edges at $\pm\theta_c(T)$. Kortman and Griffiths [3] pointed out that these two edges are singularity points, which were later termed Yang-Lee edge singularities [4]. Below the critical temperature $T_c$, the zero circles will infinitely approach to the positive real axis ($\theta_c \to 0$). The Lee-Yang theorem applies to general ferromagnetic Ising models and was later generalized to ferromagnetic Ising models of arbitrarily high spin [5-7] and other interesting types of interactions [8-10]. For general many-body systems, the zeros of the partition functions may not be distributed along a unit circle but otherwise present similar features as in the ferromagnetic Ising models.

The imaginary Lee-Yang zeros, however, have never been observed since they would occur only at imaginary magnetic field or imaginary temperature, which are not physical. In theoretical physics, a mathematical technique called Wick rotation has been employed to relate the imaginary inverse temperature to time. Therefore it is conceivable that the imaginary Lee-Yang zeros may be observed in the time domain.

Another intriguing work is the recent discovery that quantum criticality that would occur at low temperature can be probed at infinitely high temperature by a quantum probe with long coherence time [11]. By studying the quantum coherence of a probe spin weakly coupled to an Ising ferromagnet, we indeed find that the Lee-Yang zeroes are mapped to the zeros of the probe spin coherence. And in the thermodynamic limit, the coherence presents sudden death at a critical time corresponding to the edge singularities if the temperature is above the critical one.

We use a probe spin-1/2 coupled to the spin system (bath) described by equation (1), with probe-bath interaction $H_I = \lambda \sigma_z \otimes \sum_j \sigma_j \equiv \lambda \sigma_z \otimes H_1$, where $\lambda$ is a small coupling constant and $\sigma_z \equiv |\uparrow\rangle\langle\uparrow| - |\downarrow\rangle\langle\downarrow|$ is the Pauli matrix of the probe spin. We note that quantum coherence of a spin has previously been used to probe quantum criticality [12,13]. The probe strength is chosen to scale with the bath size as $\lambda \sim 1/\sqrt{N}$ which is $\ll 1$ for a large bath so that the bath is only weakly perturbed by the probe. If the probe spin is initially prepared in a superposition state as $|\uparrow\rangle + |\downarrow\rangle$ and the bath is at temperature $T$, the coherence of the probe spin, defined as $L(t) = \text{Tr}\left[e^{-\beta H(h)} e^{i(H+H_I)t} |\uparrow\rangle\langle\downarrow| e^{-i(H+H_I)t}\right] / Z(\beta, h)$, has an intriguing form as

$$L(t) = \frac{e^{-2iN\lambda t} \prod_{n=1}^{N}(e^{-2\beta h} e^{4i\lambda t} - z_n)}{\prod_{n=1}^{N}(e^{-2\beta h} - z_n)} = e^{-2iN\lambda t} \frac{Z(\beta - i2t\lambda/h, h)}{Z(\beta, h)}. \quad (2)$$

The denominator in the equation above is nonzero for real field and temperature. The numerator resembles the form of partition function with a complex inverse temperature $\beta - 2it\lambda/h$. The probe spin coherence becomes zero whenever $z' \equiv \exp(4i\lambda t - 2\beta h)$ reaches a Lee-Yang zero. Particularly for Ising ferromagnets, the Lee-Yang zeros all lie on the unit circle and therefore are mapped to the probe coherence zeros ($t_n$) for vanishing external field ($h=0$), with the correspondence relation $\exp(4i\lambda t_n) = z_n$ or $t_n = \theta_n/(4\lambda)$. The probe spin coherence in a ferromagnetic Ising bath can be written as

$$L(t) = e^{-2iN\lambda t} \prod_{n=1}^{N} \left(e^{4i\lambda t} - e^{i\theta_n}\right) / \prod_{n=1}^{N} \left(1 - e^{i\theta_n}\right). \quad (3)$$

For a bath of a finite number ($N$) of spins, the probe spin coherence will become zero for $N$ times before the coherence revival at $t = 2\pi/(4\lambda)$. At the thermodynamic limit ($N \to \infty$), the Lee-Yang zeros form a continuum cut in the complex plane, the probe spin coherence will be constantly zero between the edge singularities $\pm\theta_c$ and present a sudden death feature at the critical time $t_c = \theta_c/(4\lambda)$.

To illustrate the above ideas, we use the one-dimensional (1D) Ising model with nearest-neighbor ferromagnetic coupling and the periodic boundary condition. The 1D Ising model can be exactly solved through the transfer matrix method [18-19]. There is no finite temperature phase transition in 1D Ising model. The Lee-Yang zeros of 1D Ising model with $N$ spins have been exactly calculated [2].

Figure 1 shows the Lee-Yang zeros and probe spin coherence for the 1D Ising model with $N=10$ spins at various temperature. At infinite temperature ($\beta = 0$), all the Lee-Yang zeros are degenerate and $z_n = -1$ (Fig. 1a). Correspondingly, the probe spin coherence has one zero at $t = \theta_1/(4\lambda) = 5\pi/2$ (Fig. 1b). As a finite-size effect, the probe spin coherence presents periodic revivals at integer multiples of $2\pi/(4\lambda)$ as can be seen from equation (3). With temperature decreasing, the Lee-Yang zeros disperse on the unit circle. As shown in Fig. 1c, the probe spin coherence has 10 zeros corresponding to the Lee-Yang zeros. As temperature approaches to zero, the Lee-Yang zeros tend to be uniformly distributed on the unit circle (Fig. 1g). From equation (3), we see that if the Lee-Yang zeros are uniformly distributed [$\theta_n \to (2n-1)\pi/N$], the probe spin coherence is fully recovered up to a phase factor whenever the time is such that $4\lambda t = n2\pi/N$, and therefore the probe spin coherence oscillates periodically (as seen in Fig. 1h). The coherence oscillation can also be understood as a consequence of ferromagnetism emerging. At the critical temperature of the 1D Ising model ($T_c = 0$), the bath can be at two degenerate states $|+1,+1,\ldots,+1\rangle$ and $|-1,-1,\ldots,-1\rangle$, in which the probe spin has two opposite precession frequencies and its coherence oscillates as $L(t) = \cos(2N_p t)$.

Figure 2 shows the Lee-Yang zeros and the probe spin coherence as the bath size is increased toward the thermodynamic limit ($N \to \infty$). With increasing $N$, the Lee-Yang zeros become denser and denser between the two edge singularities. The probe spin coherence becomes almost constantly zero after the critical time corresponding to the first edge singularity ($t_c = \theta_c/(4\lambda)$). At the thermodynamic limit, the coherence presents a sudden death at the critical time. The probe spin coherence is not smooth at this point, though it is an analytical function by definition for any finite-size system. Note that such coherence sudden death, being a phase transition in the time domain, is fundamentally different from the previously discovered entanglement sudden death [14,15], which is instead caused by the non-analyticity in definition of entanglement. The coherence sudden death may provide hints to understand the quantum-classical crossover at the boundary between the microscopic and macroscopic worlds.

We now study how the coherence sudden death changes with the temperature decreasing, in particular, toward the critical temperature (which is zero in the 1D Ising model). As shown in Fig. 3 plot for an Ising model of 500 spins at various temperature, the gap between the Lee-Yang edges singularities tends to close as temperature decreases toward zero. At high temperatures, the probe spin coherence displays a sudden death as expected. The sudden death occurs at earlier time with temperature decreasing, in consistent with the smaller gap between the Lee-Yang edge singularities. When the temperature is close to the critical one (Figs. 1g & 1h), the Lee-Yang zeros become almost uniformly distributed (as more clearly seen in Fig. 1g for a smaller Ising bath) and the probe spin coherence displays pronounced oscillations. For larger baths, the oscillation features would appear for temperature closer to the critical one. At the thermodynamic limit, only at the critical temperature would the sudden death change to the oscillation feature.

We further study how the Lee-Yang zeros appear in probe spin coherence below the critical temperature by considering the two-dimensional (2D) Ising model which has a finite temperature phase transition. We consider a 2D Ising model in a square lattice with nearest neighbour coupling $J$ and the periodic boundary condition. This model under zero field is exactly solvable [16] and has a finite-temperature phase transition at $\beta J \simeq 0.44$. Figure 4 plots the Lee-Yang zeros and the probe spin coherence in a 2D Ising model of $8\times 50$ spins for various temperature. Above the critical temperature ($\beta$=0.1 and 0.4), the Lee-Yang zeroes have a gap across the positive real axis. The corresponding probe spin coherence shows a well-developed sudden-death feature (though not strictly non-analytical due to the finite size of the bath). Below the critical temperature ($\beta$=0.5 and 0.8), the Lee-Yang zeros are almost uniformly distributed along the unit circle and the probe spin coherence oscillates coherently with a period $2\pi/(4N\lambda)$, as a signature of the ferromagnetic phase.

The Lee-Yang theorem, and hence the discoveries presented here, apply to general physical systems. In particular for ferromagnetic Ising models, the unit-circle theorem holds regardless of the interaction range, geometry configurations, disorders, and dimensionality. Such universality offers a great deal of feasibility and flexibility for experimental observation of the imaginary Lee-Yang zeros. For other systems (e.g., anti-ferromagnetic Ising models), the Lee-Yang zeros may not lie on a unit circle. But one can apply an external field $h$, and get all the zeros of modulus $\exp(-2\beta h)$ according to equation (2). With the Lee-Yang zeros determined by probe coherence measurement, the partition function of an interacting many-body system can be reconstructed, from which the physical properties can be straightforwardly calculated. Thus measuring quantum coherence of a single spin (or a qubit in terminology of quantum computing) provides a *universal* tool to solve the formidable problem of interacting many-body systems, without requiring building a many-qubit quantum computer to simulate the physical systems.

## METHODS SUMMARY

The 1D Ising spin model can be exactly diagonalized by the transfer matrix method [17,18]. The evaluation of the partition function after the transformation becomes a trivial problem of diagonalization of a 2x2 matrix. The probe spin coherence is similarly calculated [which has been formulated in terms of partition functions in equation (2)]. Similarly, by the transfer matrix method, the 2D Ising model without magnetic field can be mapped to a 1D Ising model with a transverse field [18]. For a 2D Ising model in finite magnetic field, it was mapped by transfer matrix method to a 1D Ising model with both longitudinal field and transverse field, which is not exactly solvable. Instead, we numerically diagonalize the mapped 1D Ising model with both longitudinal field and transverse field. Therefore the partition function and hence the probe spin coherence for the 2D Ising model in finite magnetic field are obtained.

**Acknowledgements** This work was supported by Hong Kong Research Grants Council/General Research Fund CUHK402410, The Chinese University of Hong Kong Focused Investments Scheme, and Hong Kong Research Grants Council/Collaborative Research Fund HKU8/CRF/11G.

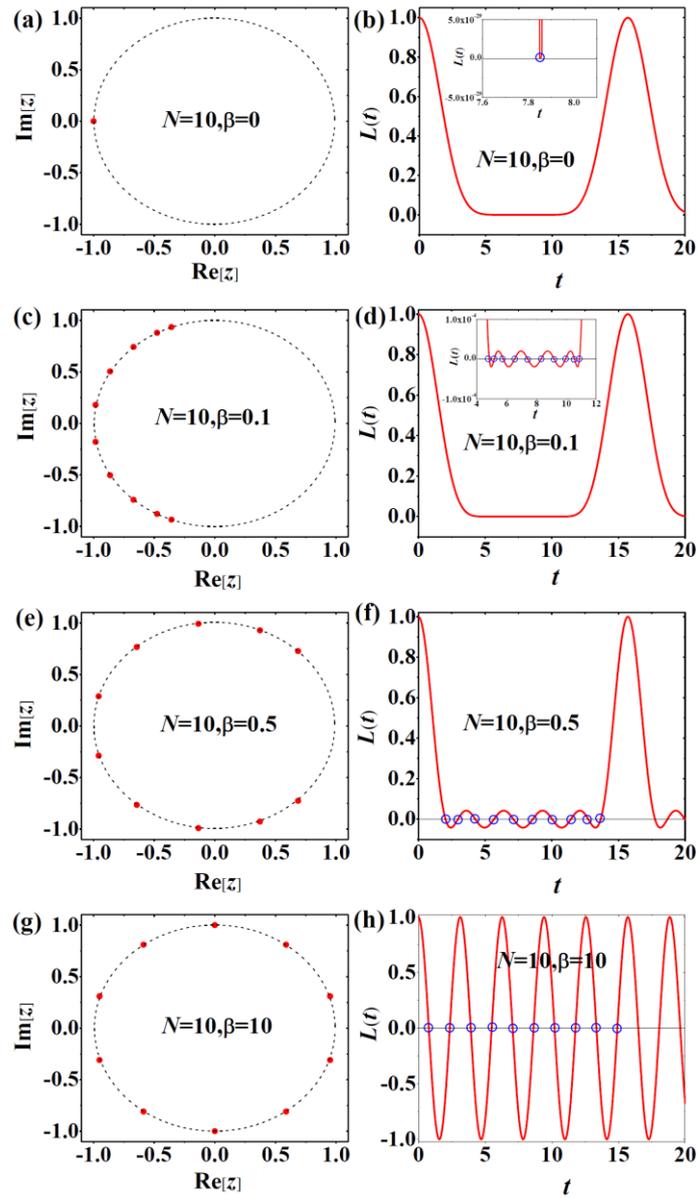

**Figure 1 | Correspondence between Lee-Yang zeros and coherence zeros of a probe spin. a**, **c**, **e** & **g**: The Lee-Yang zeros (red symbols) in the complex plane of z for a 1D Ising model of 10 spins at inverse temperature $\beta$ = 0, 0.1, 0.5 & 10, correspondingly. The black dashed lines are the unit circle. **b**, **d**, **f** & **h**: Probe spin coherence as a function of time corresponding to **a**, **c**, **e** & **g**. The insets in **b** & **d** zoom in the coherence zeros. The blue small circles in **b**, **d**, **f** & **h** mark the coherence zeros. The probe-bath coupling is such that $N\lambda^2 = 0.1$.

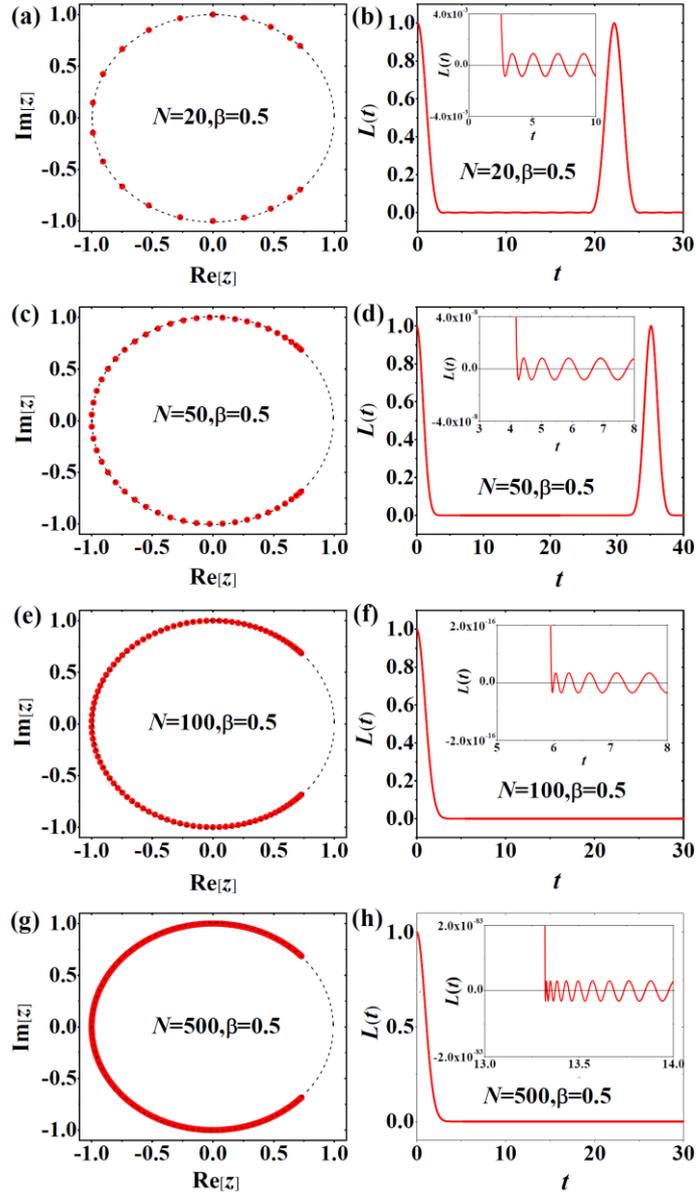

**Figure 2 | Lee-Yang edge singularities and probe spin coherence sudden death. a**, **c**, **e** & **g**: The Lee-Yang zeros in the complex plane of $z$ for a 1D Ising model with the number of spins $N$ = 20, 50, 100 & 500, correspondingly. As the thermodynamic limit is approached, the zeros form a continuum cut in the complex plane ended by two singularity points $\theta_c^{\pm}$. **b**, **d**, **f** & **h**: Probe spin coherence as a function of time corresponding to **a**, **c**, **e** & **g**. The insets zoom in the sudden death point. The parameters are such that $\beta = 0.5$ and $N\lambda^2 = 0.1$.

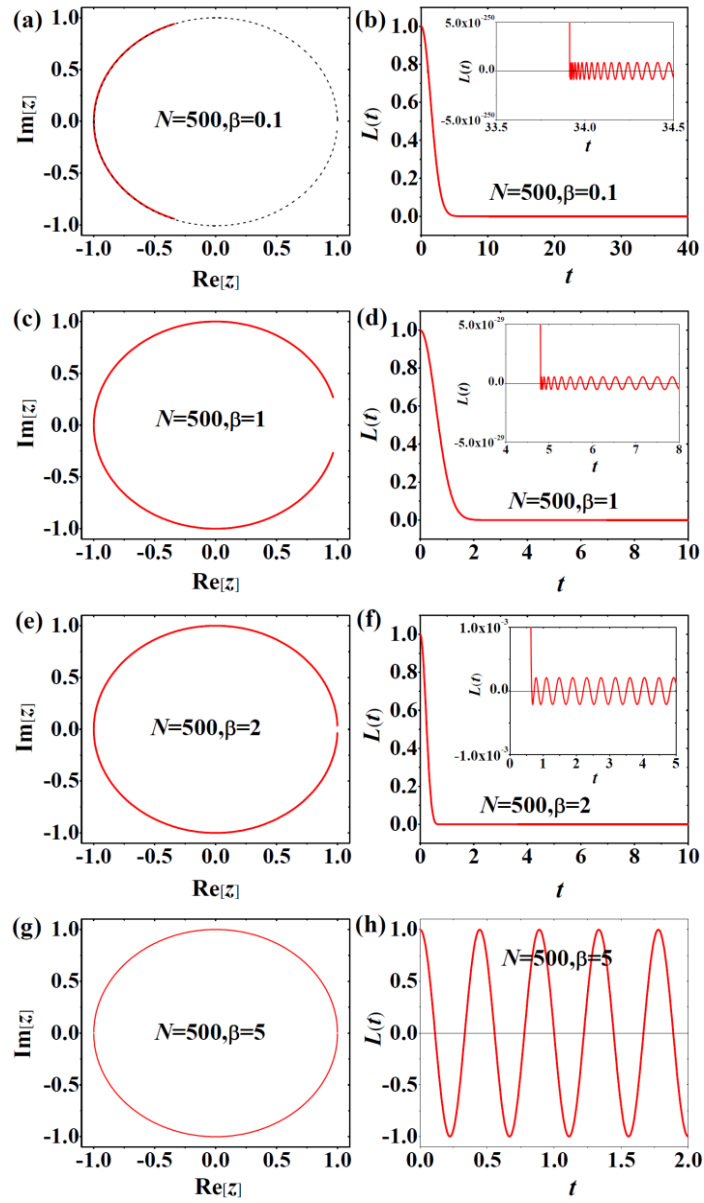

**Figure 3 | Temperature dependence of Lee-Yang edge singularities and probe spin coherence sudden death. a**, **c**, **e** & **g**: Lee-Yang zeros in the complex plane of *z* for a 1D Ising model of 500 spins at inverse temperature $\beta$ = 0.1, 1, 2 & 5, correspondingly. **b**, **d**, **f** & **h**: Probe spin coherence as a function of time corresponding to **a**, **c**, **e** & **g**. The insets of **b**, **d** & **f** zoom in the sudden death point. The probe-bath coupling is such that $N\lambda^2 = 0.1$.

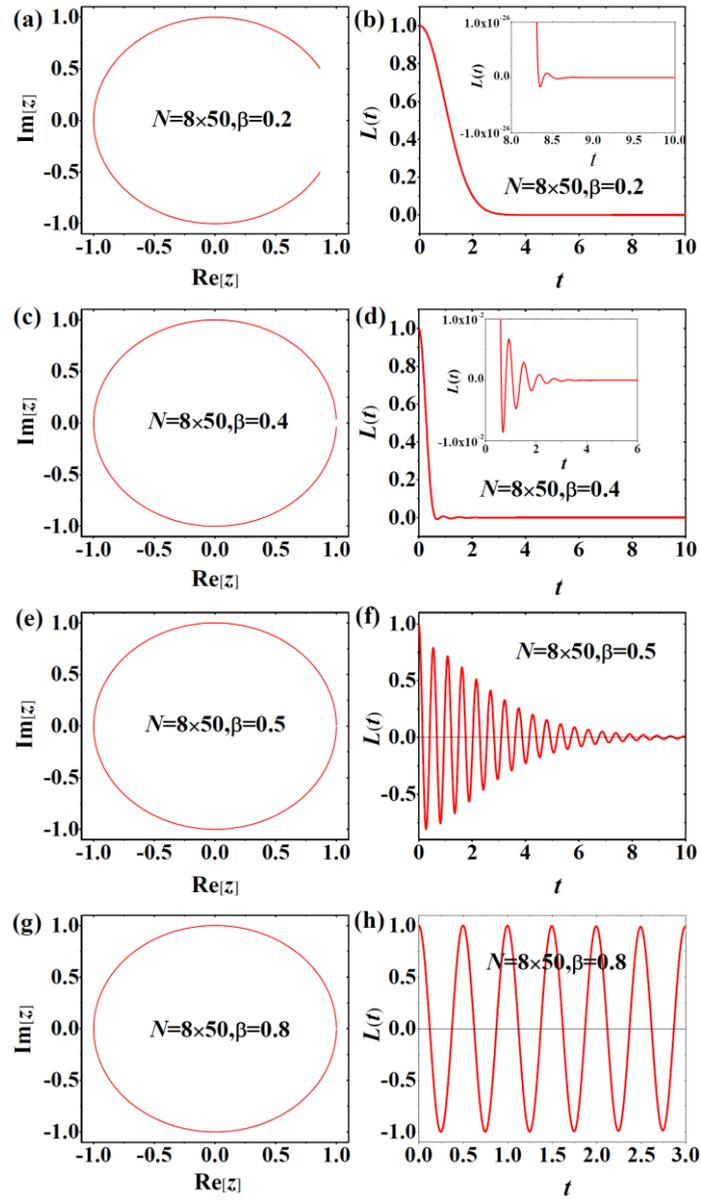

**Figure 4 | Lee-Yang zeros and probe spin coherence in a 2D Ising model at various temperature. a**, **c**, **e** & **g**: Lee-Yang zeros in the complex plane of $z$ for a 2D Ising model in a 8x50 lattice at inverse temperature $\beta$=0.2, 0.4, 0.5 & 0.8, correspondingly. **b**, **d**, **f** & **h**: Probe spin coherence as a function of time $t$ corresponding to **a**, **c**, **e** & **g**. The insets of **b** & **d** zoom in the sudden death region. The probe-bath coupling is such that $N\lambda^2 = 0.1$.